# A new introductory quantum mechanics curriculum

Antje Kohnle[1], Inna Bozhinova[1], Dan Browne[2], Mark Everitt[3], Aleksejs Fomins[1,7], Pieter Kok[4], Gytis Kulaitis[1], Martynas Prokopas[1], Derek Raine[5] and Elizabeth Swinbank[6]

[1] School of Physics and Astronomy, University of St Andrews, North Haugh, St Andrews, KY16 9SS, United Kingdom
[2] Department of Physics and Astronomy, University College London, Gower Street, London, WC1E 6BT, United Kingdom
[3] Department of Physics, Loughborough University, Loughborough, LE11 3TU, United Kingdom
[4] Department of Physics and Astronomy, University of Sheffield, Hounsfield Road, Sheffield S3 7RH, United Kingdom
[5] Department of Physics and Astronomy, University of Leicester, University Road, Leicester, LE1 7RH, United Kingdom
[6] Department of Education, University of York, Heslington, York, YO10 5DD, United Kingdom
[7] ETH Zürich, Wolfgang-Pauli-Str. 27, 8093 Zürich, Switzerland

E-mail: ak81@st-andrews.ac.uk

**Abstract**
The Institute of Physics New Quantum Curriculum consists of freely available online learning and teaching materials (quantumphysics.iop.org) for a first course in university quantum mechanics starting from two-level systems. This approach immediately immerses students in inherently quantum mechanical aspects by focusing on experiments that have no classical explanation. It allows from the start a discussion of interpretive aspects of quantum mechanics and quantum information theory. This article gives an overview of the resources available at the IOP website. The core text is presented as around 80 articles co-authored by leading experts that are arranged in themes and can be used flexibly to provide a range of alternative approaches. Many of the articles include interactive simulations with accompanying activities and problem sets that can be explored by students to enhance their understanding. Much of the linear algebra needed for this approach is part of the resource. Solutions to activities are available to instructors. The resources can be used in a variety of ways from supplements to existing courses to a complete programme.

## 1. Introduction

Many introductory university-level quantum mechanics courses and textbooks develop the theory using continuous systems (the wave mechanics approach) by introducing the Schrödinger equation and using it to find bound state and scattering solutions for a range of different potential energies. Many studies have documented student difficulties with quantum mechanics concepts in this traditional curriculum, as well as elucidating underlying reasons for these difficulties [1-15]. The wave mechanics approach may lead to incorrect ideas due to false analogies with classical systems. Examples are quantum particles losing energy when tunnelling through a potential barrier or the amplitude of the wave function being related to energy (which of course it is for a classical wave) [9, 10]. Student interest in quantum mechanics has also been shown to decrease following traditional introductory instruction [16].

Developing introductory quantum mechanics with two-level systems (two-level atoms, spin ½ particles, interferometers, qubits) has multiple advantages:

- It immediately immerses students in inherently quantum mechanical aspects of physics (complementarity, incompatible observables, single photon interference) by focusing on experiments that have no classical explanation.
- It allows a direct discussion of interpretive aspects of quantum mechanics (EPR paradox, entanglement, completeness of quantum theory, local hidden variables).
- It allows an inclusion from the start of aspects of quantum information theory (quantum key distribution, teleportation, quantum computing).
- It is mathematically less challenging, requiring only basic linear algebra such as the manipulation of 2×2 matrices instead of solving integrals and differential equations.

Incorporating quantum mechanics of two-level systems such as single photon interference experiments, entanglement of spin ½ particle pairs and the discussion of local hidden variable theories has been shown to increase student interest, and transition students away from the classical perspectives that tend to be promoted by more traditional presentations of the subject [17].

The Institute of Physics (IOP) New Quantum Curriculum provides freely available online learning and teaching materials for a contemporary approach to a first university course in quantum mechanics starting from simple two-level systems. The texts have been written by experts in the areas of quantum information (PK and DB) and foundations of quantum mechanics (ME). The materials include interactive simulations with accompanying activities developed by one of us (AK), building on the expertise of the QuVis project [18, 19]. Simulation coding was performed by four undergraduate physics students (IB, AF, GK and MP) who were also involved in the interface design and content from a student perspective. All materials have gone through a rigorous editorial process (by DR and ES). Optimization and initial evaluation of simulations and activities with students has taken place.

In what follows, we describe the New Quantum Curriculum materials in more detail. Sections 2 to 4 give overviews of the curriculum content, simulations and website structure respectively. Section 5 summarizes the work to date and describes future plans.

## 2. Overview of the curriculum content

The content of the course consists of around 80 short (typically 750 words) articles that each addresses a specific question about quantum mechanics and/or quantum mechanical systems. The articles can be read in a variety of orders; we identified five different pathways associated with different themes or approaches to the material. Students or instructors can choose a particular theme on the home page. This automatically arranges the articles for that theme in an appropriate order. The different themes are "physical", with emphasis on the physics of quantum mechanics; "mathematical", with emphasis on the mathematical structure of quantum mechanics; "historical", with emphasis on the origin of ideas in quantum mechanics; "informational", with emphasis on the theory of quantum mechanics as a way to organize information about the world; and "philosophical", with emphasis on the deep foundational issues that arise in quantum mechanics. Most articles contain a section on further reading, and a list of prerequisite articles that can be used by the students or instructor to set out their own path through the material. Many articles include an exercise for the reader. There are 17 interactive simulations with accompanying activities the students can engage with to gain a deeper understanding of the material. In addition, when people reach an article via an Internet search, it is easy to identify articles in which prerequisite concepts are explained.

The resources are aimed at students in their first year of a physics degree at a UK university, and more widely at all students studying introductory quantum mechanics and instructors teaching at this level. Given its flexibility, the resource can also support students studying more advanced quantum mechanics and can be useful to all instructors teaching this area. The resources focus on the introductory level and thus the needed prerequisite mathematics is limited. The curriculum introduces the basics of complex numbers, matrix multiplication and eigenvalue problems for two-dimensional systems and Dirac notation in an ad hoc way, so as not to overload the student with a large amount of mathematics up front. The material can thus be used for self-study by anyone interested in learning quantum mechanics with a high-school level of mathematical understanding.

The underlying philosophy of our approach to the curriculum is to present quantum mechanics as a method of reasoning about physical systems that is based on a few *Gedanken* experiments. Here we describe the approach under the informational theme. First, we consider a photon in a Mach-Zehnder (MZ) interferometer, with and without quantum non-demolition (QND) detectors in the two arms. In a regular MZ interferometer without QND detectors the photon will always trigger the same detector in the output due to constructive and destructive interference. Gaining knowledge about the path of the photon via the QND detectors destroys this interference, and the photon will trigger each detector in the output modes randomly and with equal probability. This leads to the conclusion that the path taken by the photon inside an MZ interferometer without QND detectors is not a property of the photon in the classical sense. On the basis of this experiment we construct a simple mathematical model based on two-dimensional vectors that allows us to calculate the detection probabilities. We introduce the complex phase in the MZ interferometer and apply the theory to gravitational wave detection.

Second, we consider a spin ½ particle in a Stern-Gerlach apparatus. We again describe the behaviour of the particle in terms of two-dimensional vectors, and we deduce that the spin of the particle (in the $z$-direction) must be described by a 2×2 matrix. The operator structure of quantum mechanics therefore arises naturally from our discussion of this relatively simple experiment.

Finally, we consider a two-level atom in an electromagnetic field. We construct the Hamiltonian for the atom, and ask how we should describe changes of the quantum state over time. This leads in only a few straightforward steps to the Schrödinger equation for a two-dimensional system. By introducing an interaction Hamiltonian for the process of absorbing and emitting radiation we give a non-trivial time evolution of an atom in the ground state. We use this simple theory to explain the workings of an atomic clock.

These three *Gedanken* experiments serve to elucidate the basic structure of quantum mechanics and bring to the fore the strange nature of the theory. Compared with the traditional approach, students are not overwhelmed with challenging integrals and differential equations, and can devote most of their attention to the concepts and content of the theory. Also, this approach explicitly emphasizes the central role of eigenvalue problems in quantum mechanics. In the traditional approach this tends to be hidden in the mathematical details.

After the three *Gedanken* experiments, the course continues with the development of the theory. We give the operator formalism and the postulates of quantum mechanics; we consider composite systems, leading to key modern concepts such as entanglement and teleportation; we describe the process of decoherence and extend the concept of the quantum state to the density matrix formalism;

we explain Heisenberg's uncertainty principle and we give some of the most current interpretations of quantum mechanics.

## 3. Overview of the simulations

Interactive simulations can help students to engage with and explore physics topics through high levels of interactivity, prompt feedback and multiple representations of physics concepts [20]. They can give students visual representations of abstract concepts and microscopic processes that cannot be directly observed. By choosing particular interactive elements and limiting their ranges, students can be implicitly guided in their exploration. Given its abstract nature and often counterintuitive results, simulations may arguably be particularly useful for the learning and teaching of quantum mechanics. Research-based interactive simulations for quantum mechanics have been developed and shown to improve student understanding [19, 21-30]. However, these simulations do not cover the topics of the New Quantum Curriculum or are not at the appropriate level.

The New Quantum Curriculum simulations make use of principles of interface design from previous studies [19-20, 31-33]. Simulations include common interactive components such as play controls, radio buttons, tick boxes and sliders and all have a similar look-and-feel. An introductory text gives background information and describes the experimental setup and graphics shown. The startup screen is kept as simple as possible to encourage exploration. The simulations depict simplified, idealized situations (such as no background light in the single photon experiments, 100% detector efficiencies, etc.) to reduce complexity and cognitive load. In addition to the controls view, each simulation includes a "Step-by-step exploration" view, that explains key points with text and animated highlighting and includes step controls. Through the text explanations, the simulations aim to be self-contained instructional tools. Simulations were created using Adobe Flash, with Mathematica being used to produce some of the graphics. Table 1 lists the topics of the 17 simulations developed so far.

Each simulation comes with an accompanying activity. Activities aim to promote guided exploration and sense-making, with scaffolding to help students progress from simpler to more complex situations. Activities aim to help students link different representations, use the simulations to compare and contrast situations, collect data and and interpret outcomes of their calculations. Each activity has an initial question asking students to freely explore the simulation and note things they have found out. Solutions to all activities are available for instructors.

**Table 1.** The 17 simulations developed so far, sorted by topics.

| | |
|---|---|
| Linear algebra | Matrix multiplication |
| | Graphical representation of eigenvectors |
| | Graphical representation of complex eigenvectors |
| Fundamental quantum mechanics concepts | The expectation value of an operator |
| | Superposition states and mixed states |
| | Uncertainty of spin measurement outcomes |
| | Incompatible observables: Spin 1 particles in successive Stern-Gerlach experiments |
| Single photon interference | Build a Mach-Zehnder interferometer |
| | Phase shifter in a Mach-Zehnder interferometer |
| | Interferometer experiments with photons, particles and waves |
| The Bloch sphere representation | Bloch sphere representation of quantum states for a spin 1/2 |
|                                   | particle                                                         |
|                                   | Time development of two-level quantum states in the Bloch sphere representation |
|                                   | Successive measurements in the Bloch sphere representation       |
| Entanglement and hidden variables | Entanglement: the nature of quantum correlations                 |
|                                   | Entangled spin ½ particle pairs versus local hidden variables    |
|                                   | Entangled spin ½ particle pairs versus an elementary hidden variable theory |
| Quantum information               | Quantum key distribution with entangled spin ½ particles         |

The simulations aim to help students make connections between multiple representations such as physical, graphical and mathematical representations. For example, the "Build a Mach-Zehnder interferometer" simulation shows photons passing through the experiment, as well as the matrices corresponding to optical elements and the photon quantum state at various points in the setup (see figure 1). Simulations make the invisible visible, e.g. by depicting single photons.

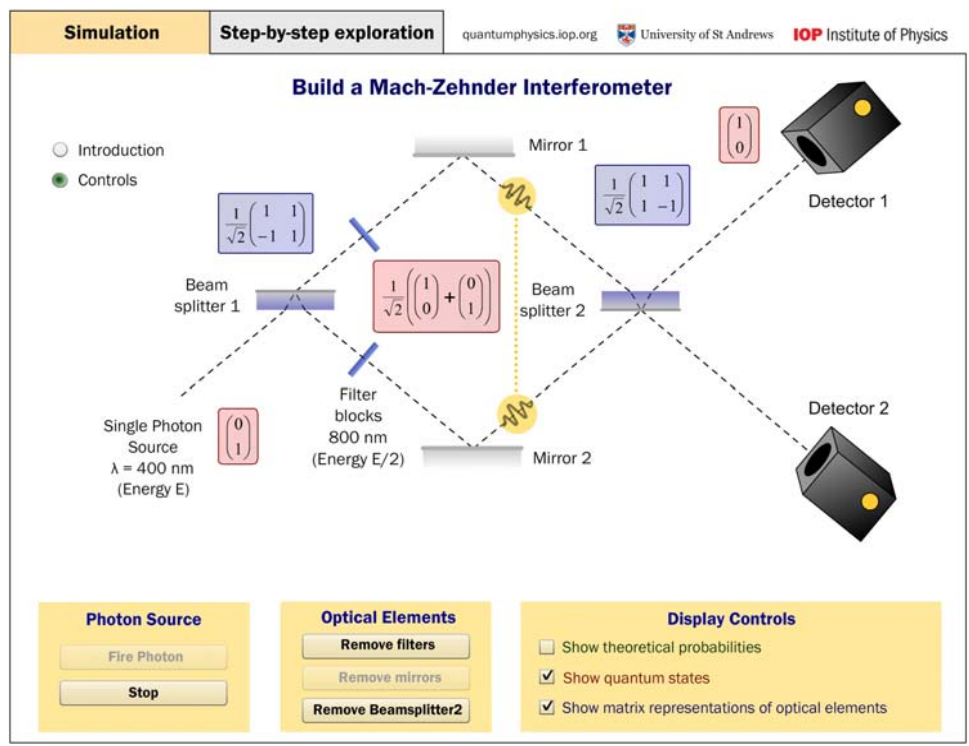

Figure 1: A screenshot of the "Build a Mach-Zehnder Interferometer" simulation.

Many simulations show users how quantum-mechanical quantities such as probabilities, expectation values and uncertainty are determined experimentally from a large number of experiments with identical inputs. Single fire, continuous fire, and fast-forward buttons allow the user to gather statistics at their chosen pace. For example, "The expectation value of an operator" simulation shows spin ½ particles all in the same input state passing through a Stern-Gerlach experiment, and shows the experimentally determined and theoretical measurement outcome probabilities and expectation value of the z-component of spin, both mathematically and graphically (see figure 2).

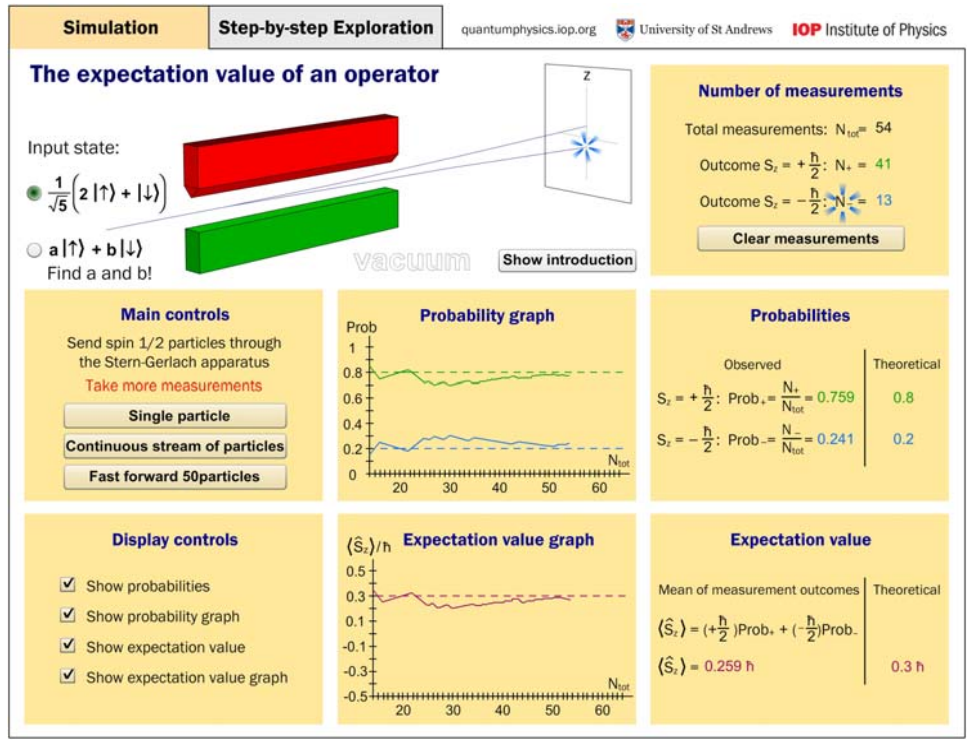

Figure 2: A screenshot of the "Expectation value of an operator" simulation.

Simulations aim to support model-building by allowing students to gather data that would be difficult or impossible to collect in reality. For example, the "Interferometer experiments with photons, particles and waves" simulation allows the user to compare and contrast the behaviour of classical particles, electromagnetic waves and photons under the same experimental conditions. Simulations aim to enhance student interpretive understanding of quantum mechanics by allowing students to test local hidden variable theories. Two simulations on hidden variables allow the user to compare the outcomes of spin measurements assuming locally pre-determined spin vectors and the actually observed outcomes predicted by quantum theory for entangled spin ½ particle pairs.

A number of simulations include small puzzles, e.g., challenges which require the student to use the simulation to find matrix elements of a transformation by determining eigenvectors and associated eigenvalues, or to determine an input state by collecting data. The second input state in figure 2 is one example of such a small puzzle. The simulations require no mathematical prerequisites apart from elementary understanding of probability and basic linear algebra with 2×2 matrices. On-demand texts and texts in the Introduction and Step-by-step Exploration explain quantities used, such as physical components of the setup, the correlation coefficient, complex exponentials, quantum state notation, etc.

We optimized simulations and activities using a total of 38 hours of observation sessions with 17 student volunteers from the appropriate introductory level at the University of St Andrews. In these sessions, students freely explored the simulations and worked through the activities while thinking aloud. Students then answered survey questions and made suggestions for improvement. We were able to trial all of the simulations and activities except one (16 in total), and had between 1 and 5 students interacting with each simulation. We also evaluated different single photon visualizations in these sessions. The lecturer of the University of St Andrews introductory quantum physics course revised the course content to include parts of the New Curriculum. We trialled three simulations and activities in this course: two in computer classroom workshops and one as a homework assignment.

We also trialled two simulations and activities as homework assignments in a modern physics course at the University of Colorado-Boulder. The observation session outcomes and analysis of homework and workshop responses led to substantial revisions in both simulations and activities that were included in all of the resources where appropriate. Further details of these evaluation efforts will be reported elsewhere.

**4. Website structure**

The website was designed to be used both on desktop and tablet computers. Due to the complexity of the simulations, these are best viewed on larger screens, with some functionality lost on a smartphone. The design is deliberately simple: all navigation is done through a panel running down the right hand side, with all content (articles, simulations, activities) in the main display. To accommodate varying screen sizes, the navigation panel can be expanded or hidden, allowing the content to be the primary focus.

When reading an article, any related articles, glossary terms and problems are displayed within the navigation panel. This allows the user to see related information while maintaining the focus on the content in the main display. To maintain the simplicity of the site, much of the additional support that IOP offers will be provided on an associated page within the main IOP website. Here, IOP will be able to provide exemplars of use, links to further resources as well as news about development and evaluation.

The resource is free to use, but all users will be asked to sign-up to the resource. Registration is to help with evaluation of the resource, to monitor use of the site and to build a community of users. To encourage new users to register, if an article is accessed via a search engine, users will be able to freely access one article before then being asked to sign up.

For instructors, the site can be used as a resource for teaching aids in their lecture courses. All content is downloadable – articles in pdf format, and simulations in shockwave – to encourage their use in lectures and as handouts. Upon request, IOP can provide all current content on the resource including solutions to individual instructors.

For students, the site allows self-directed navigation. There are many links between articles and an extensive glossary to aid learning. We have also included an option for users to record their comprehension of the site using a "traffic-light" system: Students can rate the perceived difficulty of an article in red if they feel they don't understand the content, amber if they understand part, and green if they consider themselves to have fully grasped all concepts. These icons are displayed throughout the navigation panel to allow students to easily see which articles they have read, their increasing understanding of the site as well as highlighting areas that they may be struggling with. In a later development of the site we would hope to allow instructors to access their students' self-assessment ratings.

**5. Future plans**

Evaluation so far has been limited to the simulations and accompanying activities and mostly to a single institution. We plan on multi-institutional studies in the coming year to assess the educational effectiveness of the resources in helping students from diverse backgrounds to learn introductory quantum mechanics. Outcomes of this evaluation will be used to further optimize the resources. We plan on extending the resources to include further simulations, e.g. on the Aspect experiments, decoherence and quantum information. We plan to produce HTML5/Javascript versions of the simulations. We will also be developing a second set of activities for the simulations that are more exploratory and collaborative in nature and promote student discussions. We plan to optimize these activities using group observation sessions where students collaboratively work with the simulations.

We wish to build up a community of users and will provide instructors with exemplars of use and a forum where they can exchange ideas and benefit from the experience of others. By monitoring website use and ratings and through feedback on the site, we wish to use student and instructor input to further develop these resources.

As future work, we will be investigating in more depth student conceptual understanding and difficulties with the New Quantum Curriculum content at the introductory level. This investigation will form the basis for the development of additional interactive engagement materials such as clicker questions and in-class activities, conceptual diagnostic survey questions to assess learning gains and further simulations that focus on common student difficulties.

**Acknowledgements**

We thank Christopher Hooley at the University of St Andrews for including parts of the New Quantum Curriculum in the 2013 quantum physics course. We thank Noah Finkelstein and Charles Baily at the University of Colorado-Boulder for trialling two simulations in the 2013 modern physics course. We thank Charles Baily and Bruce Torrance for carrying out a substantial fraction of the observation sessions to evaluate the simulations and activities. We thank the Institute of Physics for funding this project and developing and maintaining the project website.